\documentclass{article}

\PassOptionsToPackage{hyphens}{url}

\usepackage{arxiv}

\usepackage[utf8]{inputenc}
\usepackage[T1]{fontenc}
\usepackage{booktabs}
\usepackage{amsmath}
\usepackage{microtype}
\usepackage{tabularx}
\usepackage[numbers]{natbib}
\usepackage{tikz}
\usetikzlibrary{arrows.meta,positioning}
\usepackage[colorlinks=true, allcolors=blue]{hyperref}
\providecommand{\doi}[1]{doi:~\href{https://doi.org/#1}{\nolinkurl{#1}}}

\usepackage{orcidlink}


\hypersetup{
  pdftitle={Toward FAIR and AI-Ready Data: An Assessment of the GlueX Experiment's Data Ecosystem at Jefferson Lab},
  pdfauthor={Anil Panta, Brad Sawatzky, Casey Morean, Dmitry Romanov, Douglas Higinbotham},
  pdfsubject={Assessment of FAIR compliance and AI readiness of experimental nuclear physics data, applied to the GlueX experiment at Jefferson Lab},
  pdfkeywords={FAIR principles, AI readiness, GlueX, Jefferson Lab, nuclear physics, data management, metadata, provenance}
}

\DeclareRobustCommand{\vS}{\tikz[baseline=-0.55ex]\fill (0,0) circle (0.55ex);}   
\DeclareRobustCommand{\vP}{\tikz[baseline=-0.55ex]{\fill (0,0) -- (90:0.55ex) arc (90:270:0.55ex) -- cycle; \draw (0,0) circle (0.55ex);}} 
\DeclareRobustCommand{\vW}{\tikz[baseline=-0.55ex]\draw (0,0) circle (0.55ex);}   
\newcommand{\vN}{--}                                                    

\title{Toward FAIR and AI-Ready Data: An Assessment of the GlueX Experiment's Data Ecosystem at Jefferson Lab}
\renewcommand{\shorttitle}{Toward FAIR and AI-Ready Data: A Case Study at JLab}
\date{}
\author{
  Anil Panta\,\orcidlink{0000-0001-6385-7712}%
  \thanks{Corresponding author: \texttt{panta@jlab.org}},
  Brad Sawatzky\,\orcidlink{0000-0002-5637-0348},
  Casey Morean\,\orcidlink{0000-0001-5588-4841},
  Dmitry Romanov, and
  Douglas Higinbotham\,\orcidlink{0000-0003-2758-6526}\\[0.5em]
  Thomas Jefferson National Accelerator Facility, Newport News, VA 23606, USA
}

\begin{document}
\maketitle

\begin{abstract}
Department of Energy programs increasingly require that experimental data be not only
preserved but usable by artificial intelligence (AI) systems, yet no accepted standard
defines when a dataset is AI-ready. We argue that data must be FAIR (Findable,
Accessible, Interoperable, Reusable) before they can be AI-ready, and that AI readiness
is a distinct property requiring its own assessment. Using the GlueX experiment in
Hall~D at Jefferson Lab as a pilot, we map the data ecosystem, assess it against FAIR,
and define an AI-readiness framework whose criteria are backed by deterministic
metrics. We also evaluate existing validation software to separate what can be
assessed with off-the-shelf tools from what requires domain-specific development. The
assessment yields a set of findings, each paired with a recommendation, some of which
are already in production or prototyped and others planned as next steps.

\end{abstract}

\section{Introduction}
\label{sec:intro}

Over more than three decades of operation, Jefferson Lab (JLab) has produced
scientific data together with the software, metadata, calibration and conditions
records, workflows, and documentation needed to interpret them, and it continues to do
so. These assets represent substantial scientific value. Realizing that value over
their lifetime requires that the data and their surrounding context be managed
systematically, including the relationships among them.

JLab experiments operate under several policy and programmatic requirements of the
U.S. Department of Energy (DOE) and the Office of Science and Technology Policy (OSTP)
concerning scientific and technical information, data management, public access, and,
increasingly, artificial intelligence (AI). These requirements have different origins
and purposes, but together they establish the expectation that scientific outputs
remain discoverable, appropriately managed, and usable beyond the context in which
they were produced. DOE Order 241.1D, \textit{Scientific and Technical Information
Management}, together with statutory and public-access requirements, provides the
foundation for these
responsibilities~\cite{doe_order_241_1d,energy_policy_act_2005,ostp2022memo,doe_public_access_plan}.
Current DOE requirements for digital research data management extend these
responsibilities to research data itself~\cite{doe_data_management}. At the
laboratory level, JLab's Office of the Chief Data Officer provides data management
guidance and has set the goal of a FAIR, AI-ready data ecosystem for the
laboratory~\cite{jlab_cdo}. DOE's Office of Scientific and Technical Information
(OSTI) provides federal infrastructure for the submission, discovery, and
dissemination of this information, including the E-Link submission system and
persistent identifiers for research outputs~\cite{osti_elink,osti_doi}.

OSTI and experiment-level data management address different layers of this process.
OSTI provides federal dissemination and persistent identification of research outputs,
but an operating experiment must still maintain the metadata, provenance,
relationships, and data organization that make its scientific outputs interpretable
and reusable. The context needed to understand a single dataset may span event data,
calibrations, run conditions, software versions, configurations, workflows, and
documentation. Managing these relationships is an operational responsibility of the
experiment and of the institutional data infrastructure that supports it.

This motivates the first assessment layer of this work, the FAIR principles, under
which data should be Findable, Accessible, Interoperable, and
Reusable~\cite{wilkinson2016fair}. FAIR offers a broadly applicable framework for
evaluating whether scientific data and metadata can be discovered, accessed,
interpreted, and reused. For an operating experiment, FAIR extends beyond a final
dataset deposited in a repository. The metadata, provenance, software, configurations,
calibrations, run conditions, and relationships required to interpret a scientific
result are themselves part of the information environment in which that dataset
exists.

Satisfying FAIR does not, by itself, establish that data are suitable for AI and
machine learning (ML). A dataset may be findable and accessible while containing
missing values, uncharacterized uncertainties, sampling biases, incomplete provenance,
or other properties that limit utilization by downstream AI applications. We
therefore treat FAIR as a necessary foundation for AI readiness rather than as a
synonym for it. AI readiness adds a second assessment layer concerned with data
quality, understandability, structural quality, scientific value, bias, governance,
and application-specific validation.

This distinction is increasingly relevant as DOE initiatives, including the Genesis
Mission and the American Science Cloud, emphasize the integration of scientific data,
computing, experimental facilities, and AI
capabilities~\cite{doe_ai_strategy,genesis_mission,amsc2025}. The resulting
requirement is not simply to make more scientific data available, but to make
scientific data sufficiently structured, characterized, and governed for reliable
machine-assisted use.

A growing body of work addresses aspects of AI readiness, including data readiness
levels~\cite{lawrence2017drl}, surveys and conceptual models of readiness
dimensions~\cite{hiniduma2025survey,wang2025aiready}, quantitative inspection
tools~\cite{hiniduma2024aidrin}, the readiness of training-data pipelines at
high-performance computing scale~\cite{brewer2025drai}, and FAIR and AI-ready
scientific datasets in particle physics~\cite{chen2022fairhiggs}. These approaches
have not, however, been applied systematically to an operating nuclear physics
experiment, in which event data, calibration and conditions information, simulation,
software, workflows, and documentation are produced by different systems on different
timescales.

This paper addresses that gap using the GlueX experiment in Hall~D at JLab as a case
study. Its contributions are fivefold. First, we map the GlueX offline data ecosystem
end to end, from data acquisition through reconstruction, analysis, and simulation,
together with the metadata services and documentation that support it. Second, we
assess this ecosystem against the FAIR principles. Third, we define a domain-agnostic
AI-readiness framework of seven categories and 26 sub-criteria, each backed by a
deterministic metric, and review three existing validation packages at the
source-code level to determine which criteria can be evaluated with existing software
and which require experiment-specific development. Fourth, we derive findings and
recommendations for modernizing nuclear physics data infrastructure and report the
status of those already in production or prototyped at JLab. Fifth, we discuss the
system controls required before AI agents can operate safely on this infrastructure.

Throughout this paper, the term \emph{data} refers to any digital artifact produced,
consumed, or required by the experiment, including raw detector output, reconstructed
and analysis-level datasets, simulation, calibration and run-condition records,
configuration, software, workflows, and documentation. A scientific result depends on
these artifacts and on the relationships among them. Making data AI-ready therefore
means making these artifacts and their relationships available together, not the
event data alone.

\section{FAIR principles}
\label{sec:fairprinciples}

The FAIR guiding principles~\cite{wilkinson2016fair} define four properties that
scientific data and metadata should exhibit so that both humans and machines can find,
access, combine, and reuse them. The principles place particular emphasis on
machine-actionability, which makes them a natural prerequisite for AI readiness. We
list the fifteen sub-principles below, because the assessment in
Section~\ref{sec:fair} refers to them by their standard labels.

\begin{description}
  \item[1. Findable]\leavevmode
  \begin{itemize}
    \item[\textbf{F1.}] (Meta)data are assigned a globally unique and persistent identifier.
    \item[\textbf{F2.}] Data are described with rich metadata.
    \item[\textbf{F3.}] Metadata clearly and explicitly include the identifier of the data they describe.
    \item[\textbf{F4.}] (Meta)data are registered or indexed in a searchable resource.
  \end{itemize}

  \item[2. Accessible]\leavevmode
  \begin{itemize}
    \item[\textbf{A1.}] (Meta)data are retrievable by their identifier using a standardized communications protocol.
    \begin{itemize}
      \item[\textbf{A1.1.}] The protocol is open, free, and universally implementable.
      \item[\textbf{A1.2.}] The protocol allows for an authentication and authorization procedure, where necessary.
    \end{itemize}
    \item[\textbf{A2.}] Metadata are accessible, even when the data are no longer available.
  \end{itemize}

  \item[3. Interoperable]\leavevmode
  \begin{itemize}
    \item[\textbf{I1.}] (Meta)data use a formal, accessible, shared, and broadly applicable language for knowledge representation.
    \item[\textbf{I2.}] (Meta)data use vocabularies that follow FAIR principles.
    \item[\textbf{I3.}] (Meta)data include qualified references to other (meta)data.
  \end{itemize}

  \item[4. Reusable]\leavevmode
  \begin{itemize}
    \item[\textbf{R1.}] (Meta)data are richly described with a plurality of accurate and relevant attributes.
    \begin{itemize}
      \item[\textbf{R1.1.}] (Meta)data are released with a clear and accessible data usage license.
      \item[\textbf{R1.2.}] (Meta)data are associated with detailed provenance.
      \item[\textbf{R1.3.}] (Meta)data meet domain-relevant community standards.
    \end{itemize}
  \end{itemize}
\end{description}

\section{The GlueX data ecosystem}
\label{sec:ecosystem}

\begin{figure}[htbp]
  \centering
  \begin{tikzpicture}[
    font=\footnotesize,
    stage/.style={draw, semithick, align=center, minimum height=8.5mm, inner xsep=6pt},
    svc/.style={draw, semithick, align=center, minimum height=8.5mm, inner xsep=5pt, fill=black!6},
    arr/.style={-{Stealth[length=2mm]}, semithick},
    node distance=9mm and 8mm
  ]
    \node[stage] (daq)  {DAQ\\EVIO};
    \node[stage, right=of daq] (rec)  {\texttt{halld\_recon}\\reconstruction};
    \node[stage, right=of rec] (rest) {REST\\HDDM};
    \node[stage, right=22mm of rest] (part) {PART\\ROOT trees};
    \node[stage, right=22mm of part] (red) {reduced\\ROOT/HDDM};
    \node[svc, below=of daq] (rcdb) {RCDB\\run conditions};
    \node[svc, below=of rec] (ccdb) {CCDB\\calibrations};
    \node[svc, below=of rest] (mcw) {MCwrapper\\simulation};
    \draw[arr] (daq) -- (rec);
    \draw[arr] (rec) -- (rest);
    \draw[arr] (rest) -- node[above, font=\scriptsize] {reaction filter} (part);
    \draw[arr] (part) -- node[above, font=\scriptsize] {DSelector} (red);
    \draw[arr] (daq) -- node[right, font=\scriptsize] {register run} (rcdb);
    \draw[arr] (ccdb) -- node[right, font=\scriptsize] {constants} (rec);
    \draw[arr] (mcw) -- node[right, font=\scriptsize] {simulated events} (rest);
  \end{tikzpicture}
  \caption{Data flow of the GlueX experiment. Raw events in EVIO format are
  reconstructed by \texttt{halld\_recon} into REST (HDDM) files, reduced by the
  reaction filter into PART trees, and further reduced by DSelector-based selections
  into final files. RCDB records run conditions during data taking, CCDB supplies
  calibration constants during reconstruction, and MCwrapper produces simulated events
  that enter the same chain.}
  \label{fig:dataflow}
\end{figure}
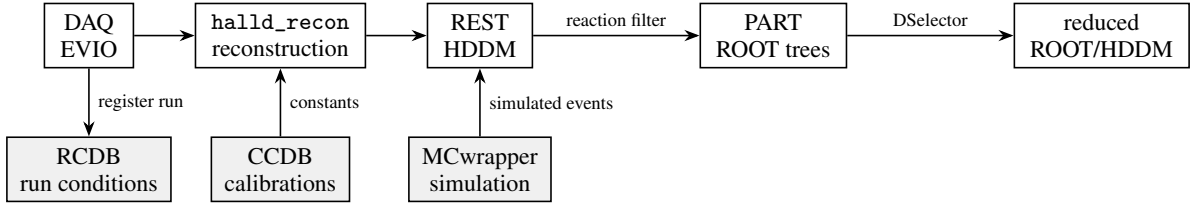

Figure~\ref{fig:dataflow} shows the GlueX processing chain and its supporting
services. The subsections that follow describe each processing stage, the metadata
services, the data formats and software, and the documentation system.

\subsection{From raw data to reconstruction}
\label{sec:raw}

Raw data are organized into \emph{runs} and written by the data-acquisition (DAQ)
system in the EVIO (EVent Input/Output) format~\cite{evio}. Each run is registered in
the Run Conditions Database (RCDB)~\cite{rcdb_github}, which records the run
identifier, timestamps, slow-control values from EPICS (Experimental Physics and
Industrial Control System)~\cite{epics1994} captured at the prestart transition, the
trigger configuration, and operator annotations. Reconstruction is performed by the
\texttt{halld\_recon} framework~\cite{halld_recon} through a configurable set of
plugins that decode detector hits and build physics objects. Its output is written in
REST (Reconstructed Event STorage), a form of the Hierarchical Data Description Model
(HDDM) format~\cite{hddm}, and contains reconstructed objects such as tracks, showers,
and vertices but no raw hit information.

Beyond the raw events, reconstruction depends on three sources of context. These are
the translation tables that map electronics channels to detector elements, the
calibration and alignment constants retrieved at run time from the Calibration
Constants Database (CCDB)~\cite{ccdb_github}, and the software version and plugin
configuration of the job. Together, the input runs, the software and its
configuration, and the resolved CCDB state define a \emph{reconstruction pass}, a
reproducible transformation from raw to reconstructed data. The pass itself, however,
is recorded only implicitly. The output path encodes the run period, data level, pass
version, file type, and run number
(\texttt{/<RunPeriod>/recon/ver<N>/REST/<run>/<file>.hddm}), so the filesystem layout
effectively serves as the metadata catalog.

\subsection{From reconstruction to analysis}
\label{sec:analysis}

Analysis proceeds in two reduction steps. First, a reaction filter processes REST
files into Physics Analysis ROOT~\cite{root1997} TTrees (PART), which are flat,
columnar event tables, applying reaction-specific preselections defined by the
requesting analyst. Batch campaigns of this step are organized as \emph{analysis
launches}. Second, analysts run selections built on DSelector~\cite{gluex_root_analysis},
the collaboration's ROOT analysis class for PART trees derived from
\texttt{TSelector}, to produce final reduced ROOT or HDDM files. The reaction
definitions, cut values, and selector configurations that define these steps are held
in analysis-specific code and wiki pages and are not captured systematically in any
queryable record.

\subsection{Monte Carlo production}
\label{sec:mc}

Simulated data are produced by MCwrapper~\cite{mcwrapper}, an orchestration layer that
accepts requests through a web form, splits them into jobs on the Open Science
Grid~\cite{pordes2007osg}, and manages generator configuration, detector simulation,
reconstruction, and software versioning. Requests, jobs, and outputs are tracked in
the MCwrapper database, and outputs, including configuration files and software
version tags, are archived to the tape library or to work disks.

Three weaknesses stand out. First, the deployed production instance was not, at the
time of assessment, traceably pinned to an identifiable release of the
version-controlled MCwrapper source. Second, submission is mediated by individual user
accounts rather than by a service identity, so request provenance depends on
operational convention. Third, event generators accept free-form inputs with no
structured schema, and environment setup is passed as shell commands embedded in a
free-text comment field.

\subsection{Metadata services}
\label{sec:metadata}

Beyond the MCwrapper tracking database (Section~\ref{sec:mc}), two databases hold the
structured metadata of the experiment.

RCDB records run-level conditions, including beam energy, event counts, and trigger
configuration, for 57 run types. This makes it the natural authority for reproducible
run selections, such as isolating production runs that satisfy specific target,
energy, and data-quality criteria. RCDB was deliberately designed as an automated
collector. Conditions are ingested from the DAQ, EPICS, and log files without human
intervention, so the record is as trustworthy as the systems that feed it. Integrity
is likewise managed at the database level, through a production instance with
per-account access rights, a read-only public replica, a staging path for occasional
manual corrections, and history preserved through multi-stage backups. This model is
appropriate for a system with no human in the loop.

Within Hall~D, the run number is the primary key of the database. It is unique across
experiments and run periods within the hall and follows a defined numbering scheme,
but it is a local identifier rather than a globally unique persistent identifier in
the FAIR sense. RCDB also stores DAQ and trigger configuration files inline, together
with their checksums. Physics data files, on the other hand, appear only as per-run
scalar conditions, namely a file count and the path of the last file, and RCDB
provides no mechanism to resolve their eventual location in mass storage. Finally,
deployments have diverged. Instances in different halls run different RCDB versions on
hall-local hardware under independent management practices, so improvements must be
redeployed for each experiment. Halls or test setups without a stable run-numbering
scheme cannot reuse the run-number-keyed design directly.

CCDB, by contrast, was designed for human-produced inputs. It stores calibration
constants, alignment parameters, and translation tables as append-only
\emph{assignments}. A new calibration never overwrites its predecessor, any past state
can be reproduced by filtering the assignment history by time, and named
\emph{variations} form a branching hierarchy in which each variation falls back to its
parent. Constants are resolved at job run time from the combination of run, variation,
and time. Because resolution depends on this context, the effective calibration state
of any past processing can be recovered only if all three are known. As calibrations
evolve, dedicated calibration workflows publish updated constants to CCDB for use in
subsequent passes.

Both databases share one design orientation, which is that they serve software. A
condition or constants table exists because some version of the reconstruction
software consumes it, and immutability is organized around software-facing
dimensions, namely run ranges, variations, and timestamps, rather than around data
products. How data products relate to one another, or to the files that carry them,
was not a design goal. This orientation underlies much of the linkage gap identified
in Section~\ref{sec:fair}.

\subsection{Data formats and software}
\label{sec:formats}

Four formats span the processing chain. EVIO holds raw data, HDDM holds structured
event records including REST, PART files are flat ROOT TTrees for analysis, and
general ROOT files hold histograms and reduced datasets. All four are formally
structured. HDDM is schema-based, and \texttt{halld\_recon} objects are typed C++
classes with documented semantics. None of the formats, however, embeds
machine-readable provenance or dataset-level metadata, so a file must be opened before
anything can be known about its content. The reconstruction software is versioned and
routinely containerized, which pins the compiler, library, and configuration state.
Downstream analysis code, including reaction filters and DSelectors, is
version-controlled in practice, but its repositories are not systematically verified
or archived.

One class of information is not recorded as data anywhere in the chain. Detector
resolutions, efficiencies, and systematic uncertainty estimates are held as the
working knowledge of detector experts, quoted as representative values in
publications, or embedded as smearing parameters in simulation code. The detector
reference paper, for example, gives a charged-particle momentum resolution of 1--3\%
and a photon energy resolution of $5$--$6\%/\sqrt{E}$ (with $E$ in GeV) in prose and
in per-subsystem figures, without a consolidated table~\cite{adhikari2021gluex}. CCDB
contains scattered per-subsystem resolution and efficiency tables. Of its 682 type
tables, 43 match resolution- or efficiency-related naming patterns, and most of these
are inputs to simulation smearing. No curated, collaboration-wide set of uncertainties
exists.

\subsection{Documentation}
\label{sec:docs}

At the time of assessment, collaboration documents were maintained in
DocDB~\cite{docdb}, a document server originally developed at Fermilab, with separate
public and private instances. DocDB
predates FAIR and modern metadata practice. Its documents carry minimal
machine-readable metadata, have no persistent identifiers, and are not linked to the
datasets or software they describe. Because AI systems need contextual documents, such
as procedures, detector descriptions, and analysis notes, to operate on experimental
data, document accessibility is part of data readiness rather than an adjacent
concern.

\section{FAIR assessment}
\label{sec:fair}

We assessed the ecosystem described in Section~\ref{sec:ecosystem} against the FAIR
principles~\cite{wilkinson2016fair} at the level of datasets and files. The assessment is a qualitative expert mapping of the ecosystem onto the fifteen
sub-principles of Section~\ref{sec:fairprinciples}. It draws on direct inspection of
the RCDB and CCDB databases, the MCwrapper deployment, and the relevant source code
repositories, on a review of collaboration documentation, and on discussions with
GlueX experimentalists, software developers, and computing staff. It reflects the state of the ecosystem at the time of assessment.
Table~\ref{tab:fair} summarizes the result, and the paragraphs below give the basis for each verdict. In summary,
GlueX metadata are substantial but \emph{unlinked}. The systems that describe the
data do not identify the data they describe.

\begin{table}[htbp]
  \caption{FAIR status of GlueX data products at the time of assessment.}
  \label{tab:fair}
  \centering
  \small
  \begin{tabular}{@{}llll@{}}
    \toprule
    \textbf{Principle} & \textbf{Sub-principle} & \textbf{Status} & \textbf{Current state} \\
    \midrule
    Findable      & F1   & Not met & No persistent identifiers for data products \\
                  & F2   & Partial & Metadata exist at run level, not product level \\
                  & F3   & Not met & Metadata do not reference data identifiers \\
                  & F4   & Not met & No searchable catalog of data products \\
    \addlinespace[2pt]
    Accessible    & A1   & Not met & Retrieval by path, not by identifier \\
                  & A1.1 & Not met & No open protocol exposed for external access \\
                  & A1.2 & Partial & UNIX permissions, not dataset-level authorization \\
                  & A2   & Partial & Metadata persist but do not name data products \\
    \addlinespace[2pt]
    Interoperable & I1   & Partial & Structure expressed in code and format schemas \\
                  & I2   & Not met & Semantic vocabulary held only in code and documents \\
                  & I3   & Not met & Processing context not recorded as provenance \\
    \addlinespace[2pt]
    Reusable      & R1   & Not met & No product-level descriptive attributes \\
                  & R1.1 & Not met & No stated usage terms \\
                  & R1.2 & Partial & Containers preserve environment; lineage informal \\
                  & R1.3 & Not met & No community metadata standard applied \\
    \bottomrule
  \end{tabular}
\end{table}

\paragraph{Findable.}
GlueX data products do not satisfy the Findable sub-principles.
\begin{description}
  \item[F1 (persistent identifier).] No globally unique, persistent identifier, such
  as a DOI, a UUID, or a catalog data identifier, is assigned to raw, reconstructed,
  or analysis-level data products.
  \item[F2 (rich metadata).] Descriptive metadata exist at the run and conditions
  level in RCDB and CCDB, but they are not attached to individual data products as
  metadata records.
  \item[F3 (metadata name the data).] Because data products have no identifiers, no
  metadata record can reference the data it describes. The association is implicit,
  through run numbers and expert knowledge.
  \item[F4 (indexed for search).] Data products are not registered or indexed in a
  searchable catalog, and discovery relies on filesystem layout and domain
  conventions.
\end{description}

\paragraph{Accessible.}
Access to GlueX data rests on filesystem conventions rather than on a retrieval
protocol. FAIR does not require data to be open, only that access conditions be
clearly defined and that retrieval follow a standard procedure. Restrictions arising
from the GlueX collaboration data policy are therefore not in themselves a FAIR gap.
The gap lies in how access is granted and how data are retrieved.
\begin{description}
  \item[A1 (retrieval by identifier).] Routine access is through POSIX filesystems on
  JLab internal systems, controlled by UNIX permissions and group membership. No
  standardized dataset-level retrieval protocol, such as HTTPS, XRootD, WebDAV, or an
  API, is offered, so retrieval presupposes knowledge of file paths and directory
  conventions.
  \item[A1.1 (open protocol).] POSIX access is not exposed through an open,
  universally implementable protocol that an external user could employ.
  \item[A1.2 (authentication and authorization).] Authorization is handled by UNIX
  permissions and group membership rather than by a dataset-level authentication and
  authorization procedure.
  \item[A2 (persistent metadata).] RCDB and CCDB metadata remain available through
  their clients and web interfaces independently of the data files, which satisfies
  the letter of this principle. Because these metadata never referenced specific data
  products, however, they cannot record exactly what was lost when files are retired.
\end{description}

\paragraph{Interoperable.}
The data formats are structured, but the vocabulary and processing context that make
events interpretable are not.
\begin{description}
  \item[I1 (knowledge representation).] The formats are formally structured, and the
  reconstruction object model is strongly typed and well specified in code. A
  calorimeter hit type, for example, carries typed fields that feed a documented
  shower-building algorithm. This structure is expressed in C++ and in
  format-specific schemas, however, rather than in a shared language for knowledge
  representation.
  \item[I2 (FAIR vocabularies).] HDDM embeds its schema in each file, but the semantic
  vocabulary of the object model, including units, definitions, and relationships,
  exists only in code and documentation rather than in a shared, FAIR-compliant
  vocabulary.
  \item[I3 (qualified references).] The processing context that gives events meaning,
  including software versions, plugin configuration, and resolved calibration state,
  is carried in unstructured configuration text, shell arguments, environment files,
  and database rows rather than in a standardized provenance representation such as
  W3C PROV~\cite{w3cprov2013}.
\end{description}

\paragraph{Reusable.}
Reuse requires collaboration membership, a JLab computing account, and
expert guidance.
\begin{description}
  \item[R1 (rich attributes).] No product-level metadata records exist, so data
  products carry no descriptive attributes to support reuse.
  \item[R1.1 (usage license).] For the same reason, no record states the usage terms
  or attribution requirements of a data product.
  \item[R1.2 (detailed provenance).] Provenance is incomplete, and reproducing a
  published selection requires locating analysis-specific code and
  collaboration-internal wiki pages. Containerized reconstruction is the strongest
  reusability asset in the chain, because it preserves the execution environment,
  although the identity of the container used for a given pass is recorded only
  informally.
  \item[R1.3 (community standards).] No domain-relevant community metadata standard is
  applied to GlueX data products.
\end{description}

\section{An AI-readiness assessment framework}
\label{sec:airead}

\subsection{Why FAIR is necessary but not sufficient}

Closing the FAIR gaps of Section~\ref{sec:fair} would not, by itself, make GlueX data
AI-ready. A dataset can be formally FAIR while critical fields are empty, provenance
is incomplete, or value distributions violate the assumptions of downstream models.
AI readiness therefore requires a second, quantitative layer, consisting of
measurable properties of the data themselves evaluated against the requirements of
the intended task~\cite{hiniduma2025survey,wang2025aiready}. Existing proposals for
this layer range from qualitative readiness levels~\cite{lawrence2017drl} to maturity
models for data preparation at high-performance computing
scale~\cite{brewer2025drai}, but none has been widely adopted.

\subsection{Assessment principles}
\label{sec:principles}

Because AI methods are probabilistic, the assessment of the data they consume should
not be. Recorded, deterministic metrics constrain the inputs to an AI analysis and
make data-related sources of error explicit. We therefore require that any readiness
assessment be
\begin{itemize}
  \item \textbf{reproducible}, meaning that identical inputs and configuration yield
  identical results, independent of the executing environment, the person performing
  the assessment, or any AI agent involved;
  \item \textbf{comparable}, meaning that results support systematic tracking of drift
  and quality across runs, run periods, and reprocessing campaigns; and
  \item \textbf{defensible}, meaning that every readiness verdict carries a documented
  justification traceable to the underlying data or to a recorded assessment rule.
\end{itemize}
These requirements exclude purely narrative or questionnaire-based assessments. Each
criterion must be backed by a deterministic metric, meaning a computable quantity or a
classification assigned by a fixed rule. For metrics that involve model fitting or
sampling, such as influence estimates or classifier-based separability, determinism is
defined relative to a recorded configuration that includes random seeds.

\subsection{Assessment categories}

We organize the assessment into seven categories comprising 26 sub-criteria, defined
in Table~\ref{tab:categories} together with the metric basis of each. The categories
draw on general machine-learning requirements, the readiness literature, and the
capabilities of existing software, and they follow most closely the taxonomy of
readiness dimensions surveyed by Hiniduma et
al.~\cite{hiniduma2025survey,hiniduma2024aidrin}. We extend that taxonomy by attaching
a deterministic metric basis to every sub-criterion and by interpreting each category
for experimental physics data.

The framework yields a profile of per-sub-criterion results rather than a single
aggregate score, because aggregation weights are task-dependent and any fixed
weighting would encode the priorities of one application into every assessment. Each
assessment must record its configuration, including the chosen outlier scores,
thresholds, and lists of expected fields, alongside its results. Reproducibility in
the sense of Section~\ref{sec:principles} is defined relative to this recorded
configuration.

The term \emph{fairness} in Category~5 requires a note. It refers to algorithmic
fairness, that is, biases built into a sample such as unequal representation across
groups, and not to the FAIR data principles. In physics data the same mathematics
applies to rare processes (class imbalance), signal and background structure (class
separability), and detector or trigger acceptance (population representation).
Several governance sub-criteria were likewise formulated for data about people.
Consent, anonymization, and demographic disclosure risk (6.1, 6.2, and 6.5) have
little bearing on detector data. We retain them so that the framework remains
applicable across domains, and note that the governance concerns most relevant to
experimental physics are collaboration data policy, embargo periods, and export
control.

\begin{table}[htbp]
  \caption{AI-readiness assessment categories, sub-criteria, and the metric basis of
  each.}
  \label{tab:categories}
  \centering
  \small
  \begin{tabularx}{\textwidth}{@{}l>{\raggedright\arraybackslash}X@{}}
    \toprule
    \textbf{Sub-criterion} & \textbf{Definition and metric basis} \\
    \midrule
    \multicolumn{2}{@{}l}{\textit{1. Quality}} \\
    1.1 Completeness & Fraction of required values that are present, computed as non-missing over expected cells \\
    1.2 Anomalies & Rate of records that deviate from the expected norm under a stated outlier score and threshold \\
    1.3 Duplication & Rate of unintended repetition, computed as one minus the fraction of unique records \\
    1.4 Data preparation practices & Documented, checkable procedures that produce the dataset from raw inputs, including declarative contracts, provenance coverage, and software versioning \\
    1.5 Timeliness & Freshness of data relative to an expected window or update cadence \\
    \addlinespace[2pt]
    \multicolumn{2}{@{}l}{\textit{2. Understandability}} \\
    2.1 Metadata availability and quality & Proportion of expected metadata fields that are present and well formed \\
    2.2 Provenance & Recorded lineage, including sources, transformations, versions, and content hashes \\
    2.3 Interfaces for data access & Mechanisms to locate, query, or retrieve the underlying dataset, distinct from the report interface of a tool \\
    \addlinespace[2pt]
    \multicolumn{2}{@{}l}{\textit{3. Structural quality}} \\
    3.1 Standardization and representation & Share of values whose runtime type matches the declared type \\
    3.2 Quality of data schema & Existence of an explicit, machine-readable schema covering types, cardinality, units, and relationships, and conformance of the data to it \\
    3.3 Infrastructure and storage & Supported file formats, storage back ends, and execution engines \\
    3.4 Data access performance & Latency and throughput of data access \\
    \addlinespace[2pt]
    \multicolumn{2}{@{}l}{\textit{4. Value}} \\
    4.1 Significance & Information a feature carries about a target quantity, measured by mutual information, predictive scores, or signal sensitivity \\
    4.2 Reference and ground truth & Availability, consistency, and correctness of reference values such as labels, calibrated measurements, and simulation truth \\
    4.3 Sample influence & Change in an aggregate result when a record is removed or perturbed, estimated by leave-one-out, influence functions, or Shapley values \\
    4.4 Data uncertainty & Per-record or per-feature uncertainties, both aleatoric and epistemic, attached to values and propagated to derived quantities \\
    \addlinespace[2pt]
    \multicolumn{2}{@{}l}{\textit{5. Fairness and bias}} \\
    5.1 Class imbalance & Disparity of class frequencies, measured by the minority-to-majority ratio or the Imbalance Degree~\cite{ortigosa2017imbalance} \\
    5.2 Class separability & Distinguishability of classes in the input space, using geometric, cluster-based, and classifier-based measures~\cite{ho2002complexity,lorena2019complexity} \\
    5.3 Group disparity & Data-level proxies for post-training fairness criteria, such as base-rate disparity and per-group sample counts \\
    5.4 Population representation & Alignment of the sample with a reference distribution, measured by drift statistics or comparison to known populations \\
    \addlinespace[2pt]
    \multicolumn{2}{@{}l}{\textit{6. Governance}} \\
    6.1 Collection & Documentation of consent, sampling, ethics, and funding context \\
    6.2 Processing and curation & Anonymization and de-identification practices \\
    6.3 Application & Usage restrictions and controls against misuse \\
    6.4 Security & Access control, encryption, audit logging, and sensitivity classification \\
    6.5 Privacy & Quantified disclosure risk, such as $k$-anonymity, $\ell$-diversity, $t$-closeness, and differential-privacy budgets \\
    \addlinespace[2pt]
    \multicolumn{2}{@{}l}{\textit{7. Application-specific}} \\
    7.1 Model-specific metrics & Performance, calibration, drift, and robustness of models built on the data \\
    \bottomrule
  \end{tabularx}
\end{table}

The categories are deliberately independent of domain and format. Even within
Hall~D, the same physics information takes several forms as it moves through the
processing chain, from raw detector hits in EVIO, to reconstructed objects in HDDM,
to flat analysis trees in ROOT, and finally to the arrays or tables that an AI
pipeline consumes. A criterion defined in terms of one format, for example a check on
the structure of HDDM records, would no longer apply once the data had been converted
to the next. Criteria defined in terms of properties of the data, such as
completeness, provenance, or uncertainty, apply at every stage. Format independence
also has two practical benefits. It matches DOE and FAIR policies, which are
themselves stated in domain-agnostic terms, and it allows general-purpose validation
software to be evaluated against the same criteria as experiment-specific tools.

The framework thus defines what an AI-readiness assessment should measure. Whether
these measurements can be made with existing software, or instead require development
specific to experimental physics, is the subject of Section~\ref{sec:tools}.

\section{Evaluating validation software against the framework}
\label{sec:tools}

\subsection{Tool selection}

The growth of AI applications has produced many tools for validating datasets and
data pipelines. Most are bound to a specific domain or application, such as
label-quality checks for medical images or drift monitoring for recommender systems,
and even the domain-agnostic tools address only part of the readiness
problem~\cite{hiniduma2025survey}. Adjacent infrastructure, such as lineage recorders,
experiment trackers, and drift monitors, records or monitors properties relevant to
readiness, including lineage, run context, and distribution drift, but does not
evaluate them as readiness metrics.

We selected three packages that represent the main paradigms in the field, namely
machine-learning pipeline assurance, scientific data stewardship, and declarative
data-quality contracts, and asked which parts of the framework each can serve.
Together they approximate what mainstream validation software delivers off the shelf,
and the remainder marks what requires custom development.

\begin{itemize}
  \item \textbf{deepchecks}~\cite{chorev2022deepchecks} is a widely used open-source
  library for validating machine-learning models and data within pipelines. Its
  operational definition of readiness is negative, in that data are considered ready
  if they will not cause silent failures during training or inference.
  \item \textbf{AIDRIN} (AI Data Readiness Inspector)~\cite{hiniduma2024aidrin} is
  aimed explicitly at scientific data readiness and is nominally the closest match to
  our framework. In practice, much of its assessment consists of formal checks on
  user-supplied descriptions of the data rather than measurements of the data
  themselves, so its effective definition of readiness is closer to FAIR-style
  stewardship.
  \item \textbf{Great Expectations}~\cite{gx2024} is a widely adopted framework for
  declarative data-quality contracts, comparable in purpose to automated verification
  systems such as Deequ~\cite{schelter2018deequ}. Its definition of readiness is the
  strictest of the three, since data are considered ready exactly when they conform to
  a programmatic contract authored by the data owner.
\end{itemize}

\subsection{Method and verdicts}

For each of the 26 sub-criteria, we reviewed each package at the source-code level,
examining the shipped checks, their algorithms, their default thresholds, and their
failure conditions. The review covered the releases of deepchecks and Great Expectations current in late
2025 and AIDRIN version 1.0 (released 26 August 2025). The verdicts reflect these
releases and may not transfer to later versions. Each combination of package and
sub-criterion received one of four verdicts. Ordered from strongest to weakest, these
are \emph{Substantive} (\vS{}), meaning a usable, well-defined implementation of the
metric; \emph{Partial} (\vP{}), meaning coverage with non-trivial caveats;
\emph{Weak} (\vW{}), meaning an indirect proxy, a formal check for the presence of
fields, or coverage only if the user writes the check; and \emph{None} (\vN{}).

Review at the level of source code matters because nominal coverage can overstate
actual capability. For metadata quality, for example, AIDRIN checks only that field
names from the DataCite~\cite{datacite2024} and DCAT~\cite{dcat2024} schemas appear as
keys in user-supplied JSON. It performs no identifier resolution, license validation,
or vocabulary check, and we therefore classify it as Weak despite the name of the
module.

\begin{table}[htbp]
  \caption{Coverage of the readiness framework by the three packages. Symbols, ordered
  from strongest to weakest, denote Substantive (\vS{}, a usable, well-defined
  implementation), Partial (\vP{}, coverage with non-trivial caveats), Weak (\vW{}, an
  indirect proxy, a field-presence check, or a user-authored check only), and None
  (\vN{}). $^{\dagger}$AIDRIN implements $k$-anonymity, $\ell$-diversity, and
  $t$-closeness correctly. These scores quantify demographic disclosure risk, which is
  not the operative privacy concern for experimental physics data, so the cell is
  rated Weak here. In domains where demographic disclosure is the concern, the
  implementation is substantive.}
  \label{tab:coverage}
  \centering
  \small
  \begin{tabular}{@{}llccc@{}}
    \toprule
    \# & Sub-criterion & deepchecks & AIDRIN & Great Expectations \\
    \midrule
    \multicolumn{5}{@{}l}{\textit{1. Quality}} \\
    1.1 & Completeness                           & \vP & \vP & \vS \\
    1.2 & Anomalies                              & \vS & \vP & \vP \\
    1.3 & Duplication                            & \vP & \vP & \vP \\
    1.4 & Data preparation practices             & \vP & \vN & \vS \\
    1.5 & Timeliness                             & \vN & \vN & \vW \\
    \addlinespace[2pt]
    \multicolumn{5}{@{}l}{\textit{2. Understandability}} \\
    2.1 & Metadata availability and quality      & \vN & \vW & \vP \\
    2.2 & Provenance                             & \vN & \vW & \vN \\
    2.3 & Interfaces for data access             & \vN & \vN & \vP \\
    \addlinespace[2pt]
    \multicolumn{5}{@{}l}{\textit{3. Structural quality}} \\
    3.1 & Standardization and representation     & \vP & \vN & \vS \\
    3.2 & Quality of data schema                 & \vN & \vN & \vS \\
    3.3 & Infrastructure and storage             & \vN & \vP & \vS \\
    3.4 & Data access performance                & \vN & \vN & \vN \\
    \addlinespace[2pt]
    \multicolumn{5}{@{}l}{\textit{4. Value}} \\
    4.1 & Significance                           & \vS & \vP & \vN \\
    4.2 & Reference and ground truth             & \vP & \vW & \vW \\
    4.3 & Sample influence                       & \vP & \vN & \vN \\
    4.4 & Data uncertainty                       & \vP & \vN & \vN \\
    \addlinespace[2pt]
    \multicolumn{5}{@{}l}{\textit{5. Fairness and bias}} \\
    5.1 & Class imbalance                        & \vP & \vS & \vW \\
    5.2 & Class separability                     & \vN & \vN & \vN \\
    5.3 & Group disparity                        & \vP & \vP & \vN \\
    5.4 & Population representation              & \vS & \vP & \vP \\
    \addlinespace[2pt]
    \multicolumn{5}{@{}l}{\textit{6. Governance}} \\
    6.1 & Collection                             & \vN & \vW & \vN \\
    6.2 & Processing and curation                & \vN & \vN & \vN \\
    6.3 & Application                            & \vN & \vW & \vN \\
    6.4 & Security                               & \vN & \vN & \vN \\
    6.5 & Privacy                                & \vN & \vW$^{\dagger}$ & \vN \\
    \addlinespace[2pt]
    \multicolumn{5}{@{}l}{\textit{7. Application-specific}} \\
    7.1 & Model-specific metrics                 & \vS & \vN & \vN \\
    \bottomrule
  \end{tabular}
\end{table}

\subsection{Results}

Table~\ref{tab:coverage} summarizes the verdicts. Of the 26 sub-criteria, 10 are
substantively covered by at least one package, and no single package substantively
covers more than five. Four further observations follow.

\emph{Strengths are disjoint.} Among the three packages, deepchecks provides
multivariate anomaly detection based on Gower distance with local outlier
probabilities, the most extensive machinery for distribution drift, combining
univariate tests with a domain-classifier score, and the only model-level metrics.
Great Expectations substantively covers schema, typing, and preparation contracts
across many storage back ends, with a configurable tolerance for each expectation.
AIDRIN contributes the strongest measure of class imbalance, the Imbalance Degree with
selectable distances~\cite{ortigosa2017imbalance}, and correct implementations of the
standard disclosure-risk scores. On inspection of the code, however, much of the
remaining nominal scope of AIDRIN reduces to checks for the presence of fields in
user-supplied metadata, simple correlations, or Boolean indicators. Its
group-disparity metric, for example, reports a Boolean flag for each group rather than
a continuous statistic. Table~\ref{tab:coverage} records these cases as Partial, Weak,
or None. Taken together, the three packages substantively cover three of the five
quality sub-criteria, three of the four structural sub-criteria, and half of the
fairness and bias sub-criteria.

\emph{Several sub-criteria lack substantive coverage.} None of the packages
substantively addresses provenance, timeliness, data access performance, security, or
curation. None measures class separability~\cite{ho2002complexity,lorena2019complexity},
sample influence beyond segment-level proxies, or per-record data uncertainty beyond
the calibration of model outputs. These gaps matter most for physics, because
uncertainty quantification and provenance are precisely where experimental data
differ from commercial tabular data.

\emph{All three packages assume flat, tabular data.} Each operates on a relational or
data-frame view, so hierarchical event data must be flattened before any check can
run. The resulting verdict then applies to a derived view rather than to the source
dataset, and properties lost in flattening, such as event structure, relations between
objects, and per-object uncertainties, are invisible to the assessment.

\emph{Default thresholds reflect other domains.} The thresholds shipped with these
packages are calibrated for industry data. The deepchecks package, for instance, fails
a dataset when the ratio of minority to majority class frequencies falls below 0.1, a
condition that is routine, and physically meaningful, in the study of rare processes.

\section{Findings and recommendations}
\label{sec:findings}

The ecosystem survey (Section~\ref{sec:ecosystem}), the FAIR assessment
(Section~\ref{sec:fair}), and the evaluation of validation software
(Section~\ref{sec:tools}) yield six findings. Findings~1--4 concern the FAIR layer,
and Findings~5 and~6 concern the AI-readiness layer. For each finding we state the
gap, the recommended remediation, and the status of that remediation as of this
writing, classified as \emph{production}, \emph{prototype}, \emph{under evaluation},
or \emph{proposed}. Table~\ref{tab:findings} summarizes the findings together with
the FAIR sub-principles or readiness sub-criteria that each addresses. The findings
are grouped by layer rather than by priority. We regard the uncertainty finding
(Finding~5) as the most valuable, for the reason given there.

\begin{table}[htbp]
  \caption{Summary of findings, the FAIR sub-principles or readiness sub-criteria each
  addresses, the remediating technology, and implementation status.}
  \label{tab:findings}
  \centering
  \small
  \begin{tabularx}{\textwidth}{@{}l>{\raggedright\arraybackslash}p{0.2\textwidth}
      >{\raggedright\arraybackslash}p{0.13\textwidth}
      >{\raggedright\arraybackslash}X>{\raggedright\arraybackslash}p{0.2\textwidth}@{}}
    \toprule
    \# & Finding & Addresses & Remediation & Status \\
    \midrule
    1 & Data products lack identities & F1, F3, F4, A1 &
      Rucio DIDs and metadata catalog & Production (EIC); GlueX in progress \\
    2 & Metadata are software-facing & F2, F3, I3 &
      DAQ auto-registration; field library; OpenAPI and MCP interfaces; expectation suites &
      Auto-registration prototyped; remainder proposed \\
    3 & Provenance not recorded & I3, R1.2 &
      Execution-time capture; Rucio--SWIF integration; PanDA &
      Prototype; PanDA under evaluation \\
    4 & Documents and software not FAIR & F1, F2, R1.1, R1.2 &
      InvenioRDM with OSTI DOI registration; Rucio linkage; FAIR4RS &
      Repository in production (DocDB migrated); OSTI integration prototyped; Rucio linkage prototype \\
    \addlinespace[2pt]
    5 & Uncertainties not recorded as data & 4.4 &
      Curated uncertainty tables in CCDB & Proposed \\
    6 & Physics-critical criteria uncovered & 1.5, 2.2, 3.4, 4.3, 4.4, 5.2 &
      Domain-native metric suite; reuse of existing packages & Proposed \\
    \bottomrule
  \end{tabularx}
\end{table}

\subsection{FAIR findings and remediation}

\subsubsection*{Finding 1: Data products have no identities}\label{sec:finding-1}

At the time of assessment, no persistent identifiers existed at any level of the data
chain, and no catalog indexed data products. Discovery therefore depended on
filesystem layout and expert knowledge.

We recommend building the identity layer on Rucio~\cite{barisits2019rucio}. Rucio
data identifiers (DIDs) are hierarchical, distinguishing files, datasets, and
containers, and this hierarchy maps naturally onto EVIO files, runs, run collections,
and processing campaigns. Run collections deserve particular attention. An identified
set of runs is precisely the object that a physics analysis consumes, and assigning it
a persistent identifier makes an analysis both citable and transportable.

This recommendation is already partly in place. Rucio is in production at JLab, where
it serves the Electron-Ion Collider (EIC) program~\cite{epic2025rucio} together with a
metadata plugin that attaches queryable metadata to each DID. Extension of the
deployment and its metadata schema to GlueX and other JLab experiments is in progress.
Neither the catalog nor the plugin is specific to any experiment, so a single
deployment can serve the entire laboratory.

\subsubsection*{Finding 2: Metadata services are software-facing, not data-facing}\label{sec:finding-2}

RCDB and CCDB serve the reconstruction software well (Section~\ref{sec:metadata}),
but they fall short of what FAIR and AI-driven uses require. Runs are not associated
with their physics data files, metadata conventions are not uniform across halls, and
no interface allows automated pipelines or AI agents to query or propose metadata
updates safely.

We recommend extending the metadata layer in four directions. The first is to link
run records to the data identities of Finding~1, replacing per-run path conditions
with resolvable references. The second is to define a facility-wide library of
standard metadata fields, such as \texttt{start\_time}, event counts, and beam
properties, so that run selections generalize across halls. The third is to provide
machine-actionable interfaces, namely an HTTP API documented with
OpenAPI~\cite{openapi2021} and an endpoint implementing an agent-oriented protocol
such as the Model Context Protocol (MCP)~\cite{mcp2024}, with scoped application-level
identities for interactive and agent actors that build on the existing database-level
access model. The fourth is to apply declarative expectation suites to RCDB
conditions and CCDB constants so that collection and calibration errors surface
immediately. Great Expectations is well suited to this flat, contract-oriented case
(Section~\ref{sec:tools}), and such suites would complement the image-based monitoring
of detector plots performed by the Hydra system~\cite{britton2024hydra}.

Of these four directions, the first has been prototyped. Data files are now
registered in the catalog directly from the DAQ, with their run association recorded
at the time of creation~\cite{swfjlab}, so that each file carries an identity from the
moment it is written. The other three directions remain proposed.

\subsubsection*{Finding 3: Provenance is not systematically recorded}\label{sec:finding-3}

The information that defines a reconstruction pass, namely its inputs, software
versions, runtime configuration, and resolved calibration state, is fragmented across
filesystem paths, command-line arguments, environment scripts, and database entries.
Analysis-level definitions, moreover, lack any structured, machine-actionable record,
which limits end-to-end reproducibility.

We recommend capturing provenance incrementally, at the point where it already exists
during execution. The orchestration layer, including launch scripts and MCwrapper,
already holds the input runs, software versions, container identities, and CCDB
resolution context of every job it submits. It can record this information as
queryable metadata on the output datasets at little marginal cost, and analysis
launches can follow the same approach. A standard vocabulary such as W3C
PROV~\cite{w3cprov2013} is a useful target representation, although the principal
value lies in capturing the context at all rather than in adopting any particular
schema. The Butler middleware of the Rubin Observatory demonstrates that catalog-driven
data management with structured provenance capture is deployable at survey
scale~\cite{lust2023rubin,jenness2022butler}.

Execution-time capture has since been demonstrated. The prototype that registers DAQ
output also records provenance at the time of registration~\cite{swfjlab}. Building on
it, the integration of Rucio with SWIF (Scientific Workflow Indefatigable
Factotum)~\cite{swif} is under development jointly with the JLab Computational
Sciences and Technology (CST) division. SWIF manages groups of related compute jobs
and their inputs and outputs, scheduling jobs with attention to data availability,
both on the JLab computing farm and at remote sites such as NERSC. Through this
integration, processing jobs consume catalog identities and register their outputs
together with provenance. In parallel, the PanDA workload management system~\cite{maeno2024panda},
which is coupled natively to Rucio in large-scale production environments, is being
evaluated through the Genesis Mission~\cite{genesis_mission} as a laboratory-wide
option.

\subsubsection*{Finding 4: Documents and analysis software are not FAIR}\label{sec:finding-4}

At the time of assessment, collaboration documents were held in DocDB, which provides
no persistent identifiers, only minimal metadata, and no links to the data and
software the documents describe, even though these documents carry the contextual
knowledge that AI systems require (Section~\ref{sec:docs}). Analysis software is
version-controlled in practice but lacks verified repositories, citable identifiers,
and machine-readable metadata (Section~\ref{sec:formats}).

We recommend treating documents and software within a single FAIR program. A
repository based on InvenioRDM~\cite{inveniordm} supports DataCite
metadata~\cite{datacite2024}, versioning, access control, and the assignment of
persistent identifiers to documents and data records. For experiment documents, we
recommend that DOIs be registered through this repository by connecting it to the
DOE identifier services operated by OSTI~\cite{osti_elink,osti_doi}, so that a
document receives its identifier as part of being deposited rather than through a
separate procedure. This connects experiment-level data management with the federal
infrastructure described in Section~\ref{sec:intro}. The FAIR for Research Software
principles~\cite{barker2022fair4rs} provide the corresponding guidance for software,
calling for verified public repositories and tagged, archived releases with persistent
identifiers, so that the software component of provenance (Finding~3) resolves to
identified artifacts. Placing documents, datasets, and software records under one
metadata scheme, an approach also pursued by European open-data initiatives through
standardized metadata~\cite{conde2024metadata}, makes contextual information linked
and retrievable.

Part of this recommendation is now in production. An InvenioRDM repository has been
deployed at JLab, and the collaboration documents previously held in DocDB have been
migrated to it. The migrated documents do not yet carry DOIs. DOIs for JLab outputs
are currently registered through a separate laboratory procedure that submits records
to OSTI, and an integration of InvenioRDM with OSTI that would issue DOIs directly
from the repository has been tested. Linking the repository to the Rucio catalog and
to the SWIF integration of Finding~3, so that documents, dataset records, and software
releases share a common identity and metadata layer, is now being prototyped.

\subsection{AI-readiness findings and remediation}

\subsubsection*{Finding 5: Uncertainties are not recorded as data}

The resolutions, efficiencies, and systematic uncertainty estimates that turn event
counts into physics results are known to collaboration members and detector experts,
reported in the text and figures of publications, and embedded as smearing constants
in simulation software (Section~\ref{sec:formats}). No single system collects them,
and locating a value therefore requires searching across these sources with no
guarantee of completeness.

A second problem concerns validity. The detector and the reconstruction software
evolve over time. Forward-calorimeter resolutions published before the lead-tungstate
insert upgrade, for example, describe a detector that has since changed. Publications
rarely state the run ranges and software versions to which their numbers apply, so
even a value that has been located cannot be confirmed to describe the data at hand.
Sub-criterion~4.4 identifies exactly this property as a readiness criterion, and none
of the evaluated tools supplies it (Table~\ref{tab:coverage}). A human-led analysis
compensates by consulting experts. An automated or AI-driven analysis cannot, and
without recorded uncertainties it will produce results whose accuracy cannot be
defended.

We recommend creating and maintaining a collaboration-curated set of uncertainty
tables in CCDB, organized by subsystem and qualified by validity intervals. The
append-only assignments, run-range validity, and branching variations of CCDB already
provide the required mechanics. Each stored value is tied to the runs and the software
variation it describes, which resolves the location problem and the validity problem
together. We consider this the highest-value recommendation of this work, because the
value of an analysis, whether automated or not, lies not in how many numbers it
produces but in the precision that can be attached to them.

This recommendation has not yet been implemented. Together with Finding~6, it is the
designated next step of this work.

\subsubsection*{Finding 6: No evaluated tool covers the physics-critical sub-criteria}

The evaluated packages address statistical quality, schema contracts, and imbalance
and disparity metrics on flattened data. They do not substantively cover provenance,
per-record uncertainty, record-level sample influence, class separability,
timeliness, or data access performance (Table~\ref{tab:coverage}).

We recommend implementing the uncovered sub-criteria as a domain-native metric suite
that operates directly on event formats such as HDDM and PART trees. Existing packages
should be reused wherever the tabular paradigm fits, for example expectation suites on
analysis trees and metadata (Finding~2) and deepchecks drift monitoring between run
periods, rather than reimplementing capabilities they already provide well. Adopting
these packages also requires recalibrating their default conditions for physics data,
not only adding adapters for physics formats.

This recommendation is likewise proposed. The metric suite, together with the
uncertainty tables of Finding~5, is a prerequisite for the first scored readiness
assessment of a GlueX dataset.

\subsection{Institutional policy}

Technical remediation alone does not make practices durable. Drawing on the findings
above, we have prepared a data policy recommendation for JLab~\cite{morean2026policy}, to be distributed to
laboratory management. It recommends that experimental data products receive
persistent identifiers and be registered in the laboratory catalog at the time of
creation, that provenance be captured by the workflow systems that produce the data,
that documents and research software be managed as citable outputs alongside the data
they describe, with DOIs for experiment documents registered with OSTI through the
laboratory repository, and that the uncertainties needed to interpret experimental
data be recorded and maintained as data. Because the supporting infrastructure,
comprising Rucio, SWIF, InvenioRDM, and potentially PanDA, is shared across
experiments, the policy can be applied across the laboratory rather than experiment by
experiment.

\section{Discussion: operational readiness for scientific agents}
\label{sec:agent-readiness}

The remediation described in Section~\ref{sec:findings} builds the infrastructure
through which data, metadata, and provenance will be accessed, and that infrastructure
is also the environment in which AI agents may act. The framework of
Section~\ref{sec:airead} evaluates whether data and their recorded context are
suitable for machine-assisted use. Scientific agents, however, do not consume a single
static dataset. One scientific question may require an agent to
resolve a run selection, locate the corresponding event data, determine the
calibrations and software state that apply to those runs, and identify the
documentation needed to interpret the result. The agent may also follow provenance
links to derived products and add them to the context of its analysis. We therefore
distinguish \emph{data readiness} from \emph{operational readiness for agents}. The
latter is a property of the surrounding system rather than of the data, and it is
evaluated by conformance to controls on interfaces, identity, provenance, workflows,
and approval. We describe these controls in terms of three planes of an agent
architecture.

\paragraph{Read and discovery plane.}
Read-only discovery should be broad within the visibility authorized for the
requester, machine-queryable, and free of side effects. An agent should be able to
search across related resources held in several systems, including data identifiers,
repository records, schemas, workflow definitions, and provenance records. Each
returned object should state its authoritative source, stable identifier, schema
version, and validity interval. An agent can then traverse and compare scientific
context without direct access to production database tables and without relying on
filesystem conventions. Retrieval of restricted content remains subject to the access
policy of the source service, even when descriptive metadata are broadly
discoverable.

\paragraph{Action plane.}
Write operations require a different trust model. A language model may select or
parameterize a typed operation, but authentication, authorization, and schema
validation are performed by deterministic application code. Interfaces documented with
OpenAPI and agent-oriented protocols such as MCP can
expose these operations, but an interface does not itself confer
authority~\cite{openapi2021,mcp2024}. Agent tools should call supported domain APIs
rather than receive raw database credentials, general shell access, or unrestricted
service tokens. A delegated write should use a short-lived credential bound to its
intended audience, and it should identify both the human on whose behalf the request
is made and the agent or service that executes it~\cite{rfc8693}. Its authority should
be limited by tool, action, target resource, time, and approval state. For
consequential changes, approval is a separate step, bound to the exact proposal and
requiring direct human confirmation.

\paragraph{Evidence plane.}
For a consequential task performed by an agent, the system should record a context
manifest and an action record. The context manifest identifies the objects,
revisions, validity times, queries, and validation reports supplied to the agent. The
action record identifies the human principal, the executing agent or service, and the
versions of the tools and policies in force. Together they extend scientific
provenance~\cite{w3cprov2013} from data production to machine-assisted
interpretation and action, and they permit later review even though natural-language
generation is nondeterministic. Retaining full prompts is neither necessary nor always
desirable. The durable requirement is enough structured evidence to reconstruct what
information was available and which deterministic controls governed the result.

Table~\ref{tab:agent-capabilities} groups common agent capabilities by the
consequence of their use and gives the system boundary that each requires.

\begin{table}[htbp]
  \caption{Agent-facing capability classes and the system boundary each requires.}
  \label{tab:agent-capabilities}
  \centering
  \small
  \begin{tabularx}{\textwidth}{@{}>{\raggedright\arraybackslash}p{0.19\textwidth}>{\raggedright\arraybackslash}X>{\raggedright\arraybackslash}X@{}}
    \toprule
    \textbf{Capability} & \textbf{Typical scientific use} & \textbf{Required boundary} \\
    \midrule
    Read and discovery & Find runs, data products, calibrations, schemas, documents, workflows, and provenance; compare versions and validity. & Policy-filtered read scope; stable source identifiers and citations; query limits; no side effects. \\
    Proposal & Draft a run-condition correction, run collection, repository record, calibration candidate, or campaign request. & Delegated user and agent identity; schema and conflict validation; immutable proposal; no implicit approval. \\
    Execution request & Stage data or invoke a registered reconstruction, simulation, calibration, or analysis workflow. & Closed input manifest; allowlisted workflow and version; bounded parameters and resources; workload identity; recorded outputs and status. \\
    Approval support & Summarize differences, validation results, uncertainties, policy blockers, or reviewer evidence. & Read-only recommendation tied to exact sources; no approval authority; no silent waiver of failed checks. \\
    High-impact or administrative action & Approve a calibration, publish a record, delete a required replica, change access, migrate a schema, or alter service policy. & Normally excluded from autonomous tools; explicit user presence, stronger authentication, separation of duties, exact proposal binding, and complete audit. \\
    \bottomrule
  \end{tabularx}
\end{table}

A further precondition is that the institution identify which workflows an agent may
invoke. The existence of a script, or of a commonly copied command, does not establish
scientific authority. A workflow exposed to agents should have an owner with
scientific authority over it, a versioned contract for its inputs and outputs, an
identified software and execution environment, bounded parameters, and a defined
provenance record. Once such a workflow is registered, an agent may resolve its
inputs, request its execution, monitor its state, and assemble validation evidence.
Workflow management systems that already track job definitions and link them to
catalog identities, such as the SWIF and PanDA integrations described in Finding~3,
provide a natural basis for such a registry.

For GlueX, this separation would allow an agent to discover a reproducible set of
runs, follow those runs to the corresponding EVIO, REST, PART, or reduced products,
resolve their calibration and reconstruction context, and retrieve the documents and
uncertainty records needed to interpret them. The same agent could propose a run
collection or a metadata correction, or request an approved processing workflow,
without acquiring authority to approve a calibration, publish a collaboration
product, change access policy, or delete managed data. Broad read-only discovery
combined with narrowly delegated action makes an agent scientifically useful without
weakening collaboration oversight.

None of these capabilities requires a bespoke agent framework. General-purpose agent
environments and orchestration libraries can implement all three planes, provided that
access to tools is mediated by the controls described above. Support for OpenID
Connect (OIDC)~\cite{oidc2014} in such tools allows the execution environment of an
agent to be scoped by a short-lived token issued on behalf of an authenticated user.

\section{Limitations}
\label{sec:limitations}

This is a single-facility case study. The criteria are domain-agnostic by
construction, but their transferability has not yet been tested at a second facility.
The coverage verdicts were assigned by the authors under the four-level rubric without
independent replication, and the FAIR verdicts likewise rest on the authors' judgment
rather than on automated indicator tests. The framework has not yet been executed end
to end on a GlueX dataset, so Table~\ref{tab:coverage} maps the capabilities of tools
rather than measured readiness scores. The coverage analysis samples three packages as
representatives of their paradigms, so a verdict that a sub-criterion is uncovered
describes the mainstream validation paradigms at the time of review, not every
existing tool. Finally, implementation status is reported as of this writing, and the
prototypes described in Section~\ref{sec:findings} have not yet been evaluated at
production scale.

\section{Conclusion}
\label{sec:conclusion}

This work set out to establish how the data of an operating nuclear physics
experiment can be made ready for AI, using GlueX as a pilot. We mapped the GlueX data
ecosystem from data acquisition through reconstruction, analysis, and simulation,
assessed it against the FAIR principles, defined an AI-readiness framework of seven
categories and 26 sub-criteria backed by deterministic metrics, and evaluated three
representative validation packages against that framework.

The FAIR result is structural. GlueX holds substantial metadata but no mechanism that
binds them to identified data products, so findability and reusability fail at the
final step. The survey adds a second structural gap, namely that the uncertainties
that give the data their physical meaning are not recorded as data at all. The
tooling result is quantitative. Of the 26 sub-criteria, 10 are substantively covered
by at least one package, and no package substantively covers more than five. Coverage
is concentrated in statistical and schema checks on tabular data, while the
sub-criteria most characteristic of experimental physics, namely provenance,
per-record uncertainty, sample influence, and class separability, are covered only by
weak or partial proxies.

These results define a single program of work. It consists of recording uncertainties
as curated data in CCDB, completing the identity and catalog layer with Rucio, making
metadata services machine-actionable and validated against declared expectations,
capturing provenance where it already exists at execution time, bringing documents and
software under FAIR practice with InvenioRDM and the FAIR4RS principles, and building
the domain-native metrics that existing tools lack. Parts of this program are already in place. The Rucio catalog and its metadata plugin
are in production at JLab, and automated registration and provenance capture from the
DAQ have been prototyped. An InvenioRDM repository has been deployed and the DocDB
collection migrated to it, DOI registration through OSTI has been tested, and the
integration of the catalog with SWIF and InvenioRDM is under development with the CST
division. The findings have also informed a data policy recommendation~\cite{morean2026policy} for the
laboratory, and we have outlined the system controls under which AI agents could
operate safely on this infrastructure. The immediate next steps are the curated
uncertainty tables, the domain-native metric suite, and a first scored readiness
assessment of a GlueX dataset.

Nothing in the framework or the findings is specific to Hall~D. The categories are
domain-agnostic by design, and the systems assessed, namely run databases, calibration
databases, filesystem catalogs, and document servers, have counterparts at other
nuclear and particle physics facilities. The infrastructure now being deployed is
likewise shared across experiments. The framework is therefore offered as a reusable
instrument. Applied to another experiment, it yields a coverage profile from which
remediation priorities follow directly. The computing infrastructure and workforce
training that laboratory-wide adoption would require remain to be assessed in future
work.

\section*{Acknowledgments}
This work was supported by the Laboratory Directed Research and Development (LDRD)
program at Thomas Jefferson National Accelerator Facility under project LD2615. We thank the members of the
GlueX Collaboration and the JLab Computational Sciences and Technology division who
contributed their time and expertise to this study. This material is based upon work supported by the U.S. Department of Energy, Office of Science, Office of Nuclear Physics under Contract No. 89243126CSC000213, under which SURATech,
LLC operates the Thomas Jefferson National Accelerator Facility.

\bibliographystyle{unsrtnat}
\bibliography{references}

\end{document}